\documentclass[3p,times,procedia]{elsarticle}
\usepackage{nupha_ecrc}
\newcommand{\kst}{$\mathrm{K}^{*0}$}
 \newcommand{\pt}{$p_\mathrm{T}$}
\newcommand{\mpt}{$\langle p_{\mathrm{T}} \rangle$}
\newcommand{\snn}{$\sqrt{s_{\mathrm{NN}}}$}

\volume{00}

\firstpage{1}

\journalname{Nuclear Physics A}

\runauth{A. Khuntia}

\jid{nupha}

\jnltitlelogo{Nuclear Physics A}

\usepackage{amssymb}

 \usepackage{lineno}

\usepackage[figuresright]{rotating}

\begin{document}

\begin{frontmatter}



\dochead{XXVIIIth International Conference on Ultrarelativistic Nucleus-Nucleus Collisions\\ (Quark Matter 2019)}

\title{Latest results on the production of hadronic resonances in ALICE at the LHC}


\author{Arvind Khuntia (For the ALICE Collaboration)}

\address{The H. Niewodniczanski Institute of Nuclear Physics, Polish Academy of Sciences, PL-31342 Krakow, Poland}

\begin{abstract}
Measurement of short-lived hadronic resonances are used to study different aspects of particle production and 
collision dynamics in pp, p--A and relativistic heavy-ion collisions. The yields of resonances are sensitive to the 
competing processes of hadron rescattering and regeneration, thus making these particles unique probes of the
properties of the late hadronic phase. Measurements of resonances with different masses and quantum numbers
also provide insight into strangeness production and processes that determine the shapes of particle momentum
spectra at intermediate transverse momenta, as well as the species dependence of hadron suppression at high
momentum.  We present the comprehensive set of results in the ALICE experiment with  unprecedented precision
for $\rho(770)^{0}$, K$^{*}(892)$, $\phi(1020)$, $\Sigma(1385)^{\pm}$, $\Lambda(1520)$, and
$\Xi(1530)^{0}$ production in pp, p--Pb, Xe--Xe and Pb--Pb collisions in the energy range $\sqrt{s_{\rm NN}}$ = 2.76-13 TeV,
including the latest measurements from LHC Run 2. The obtained results are used to study the system-size and 
collision-energy evolution of transverse momentum spectra, particle ratios and nuclear modification factors and to
search for the onset of collectivity in small collision systems.  We compare these results to lower energy 
measurements and model calculations where available.
\end{abstract}

\end{frontmatter}
\section{Introduction}
\label{QM:intro}

At the LHC, measurements of various observables in p--Pb and pp collisions with high charged-particle multiplicity have shown remarkable similarities to corresponding measurements in heavy-ion collisions. The observation of azimuthal correlations amongst particles and anisotropic flow ($V_2$) in small collision systems indicate presence of collective effects.
However, the origins of such effects in smaller collision systems lack explanation leaving the question whether the underlying causes are the same as in large 
collision systems such as Xe--Xe and Pb--Pb. Recent results show an enhancement in the production of strange hadrons; the ratio of yield of strange hadrons to non-strange hadrons show an increase with increasing charged particle multiplicity in pp collisions and  the values of ratios in high multiplicity collisions approach the observed ratios in p--Pb and peripheral Pb--Pb collisions at similar multiplicities \cite{alicenature}.
The strength of this enhancement increases with increasing strange quark content of the hadron rather than mass or  
baryon number  in small systems. The $\phi$ meson with net strangeness zero, can help to understand strangeness production in smaller collision systems.
Comparison of  $\phi$-meson production relative to other hadrons with different strangeness content as a function of charged particle
multiplicity in various collision systems may reveal information about the effective strangeness of the $\phi$ meson. 
Furthermore, resonances with short lifetimes are sensitive to the rescattering and regeneration process during the evolution of the fireball from
chemical to kinetic freeze-out. An observable such as the nuclear modification factors of the resonances can help to understand the in-medium
 parton energy loss in heavy-ion collisions, and hence has paramount importance in understanding the particle production mechanism in high energy collisons.

 \section{Analysis details}
\label{QM:analysis}
The ALICE detector is described in detail in ref. \cite{Abelev:2014ffa}. The sub-detectors which are relevant to 
this analysis are the Time Projection Chamber (TPC), the Time-of-Flight detector (TOF), the Inner Tracking System (ITS) 
($|\eta| < 0.9$), and the V0A ($2.8 < \eta < 5.1$) and V0C ($-3.7 < \eta < -1.7$) detectors. The TPC and ITS are used for 
tracking and finding the primary vertex,  whereas the TPC and TOF are used for particle identification. The V0 detectors are used for
triggering and estimation of multiplicity at forward rapidities \cite{vocen}. The measurements of resonance, strange and multi-strange 
hadron production are carried out at mid-rapidity ($|y| < 0.5$ in pp, Xe--Xe and Pb--Pb collisions and $0 < y_{\rm{cm}} < 0.5$ in p--Pb collisions) as a 
function of the charged particle multiplicity. Resonances are measured via an invariant mass analysis from the hadronic decay daughters: 
$\rho^0 \rightarrow \pi^+\pi^- (\rm{B.R.} \sim 100 \%)$, \kst $\rightarrow$K$^{+}\pi^{-}$ (66.6 \%), 
$\phi\rightarrow$K$^{+}$K$^{-}$ (49.2 \%), $\Lambda \rightarrow$ p$\pi^{-}$ (63.9\%), $\Sigma^{*+} (\Sigma^{*-})\rightarrow  \Lambda\pi^{+}( \Lambda\pi^{-})$
(87\%) and $\Xi ^{*0}\rightarrow \Xi^{-}\pi^{+} (66.7 \%)$ \cite{Tanabashi:2018oca}. In the invariant-mass method  the combinatorial background has been estimated by using 
an event-mixing technique for resonances, however a  like-charge background is also used in some analyses. For strange and multi-strange hadrons, a set of topological cuts is applied to eliminate the background which
do not fit the expected decay topology. The raw yields are extracted from the signal distribution after the subtraction of the combinatorial background 
and are corrected for the detector acceptance, tracking efficiency and branching ratio.
\section{Results and discussion}
\label{QM:results}
\begin{figure}[ht!]
\begin{minipage}[c]{0.65\textwidth}
\vspace{-0.3 cm}
\centerline{
\includegraphics[scale=0.40]{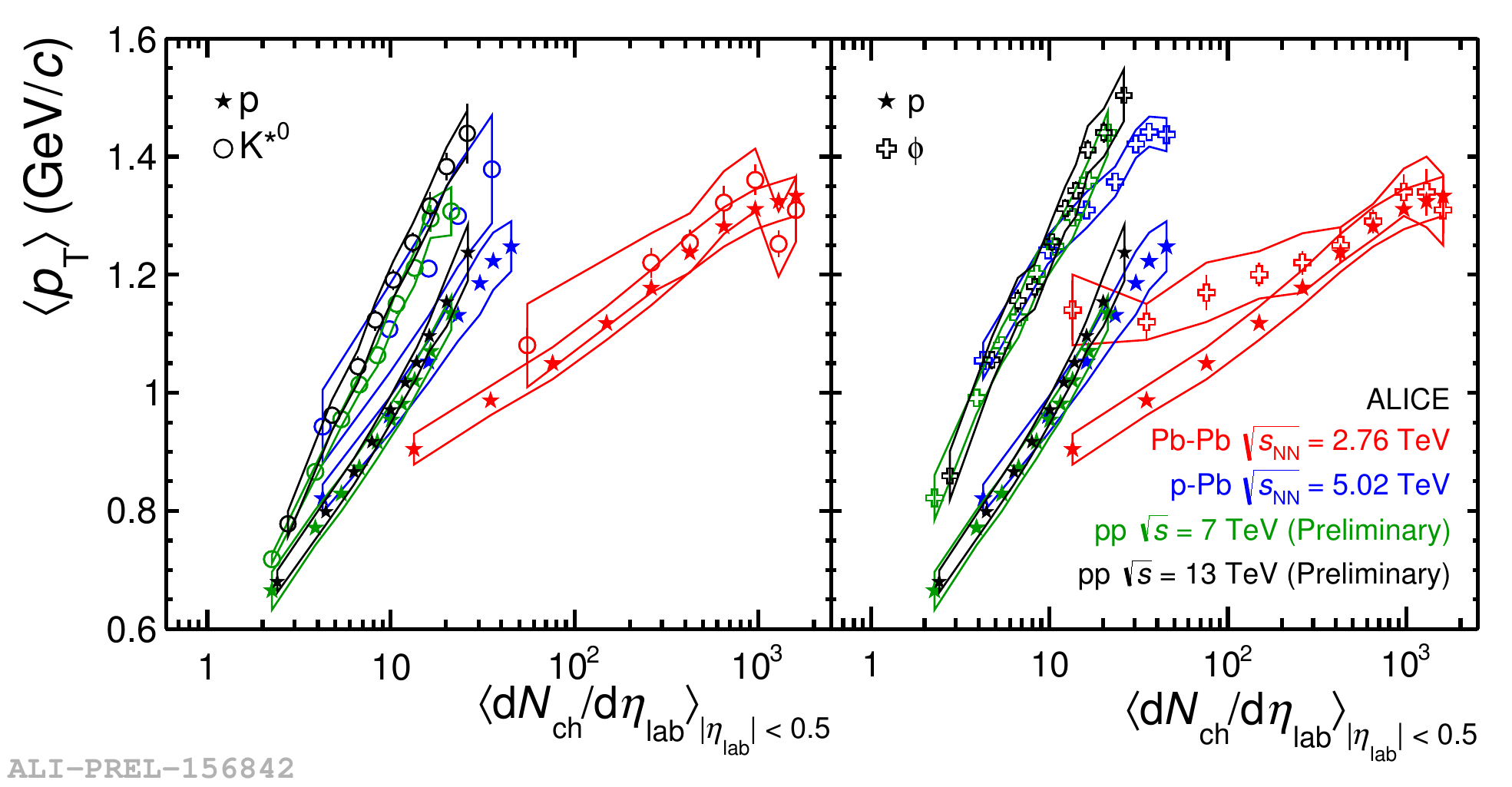}
}
\vspace{-0.3cm}
\end{minipage}\hfill
\begin{minipage}[c]{0.3\textwidth}
\caption{(Color online) Mean transverse momentum (\mpt)~ of \kst, $\phi$ and p in pp at $\sqrt{s}$ = 7 and 13 TeV, p--Pb 
at \snn~= 5.02 TeV and Pb--Pb at \snn~= ~2.76 TeV collisions as functions of charged particle multiplicity. The bars and lines represent the statistical 
and the systematic errors, respectively.}
\label{QM:meanpt}
\end{minipage}
\end{figure}

\begin{figure}[ht!]
\begin{minipage}[c]{0.65\textwidth}
\vspace{-0.3 cm}
\centerline{
\includegraphics[scale=0.23]{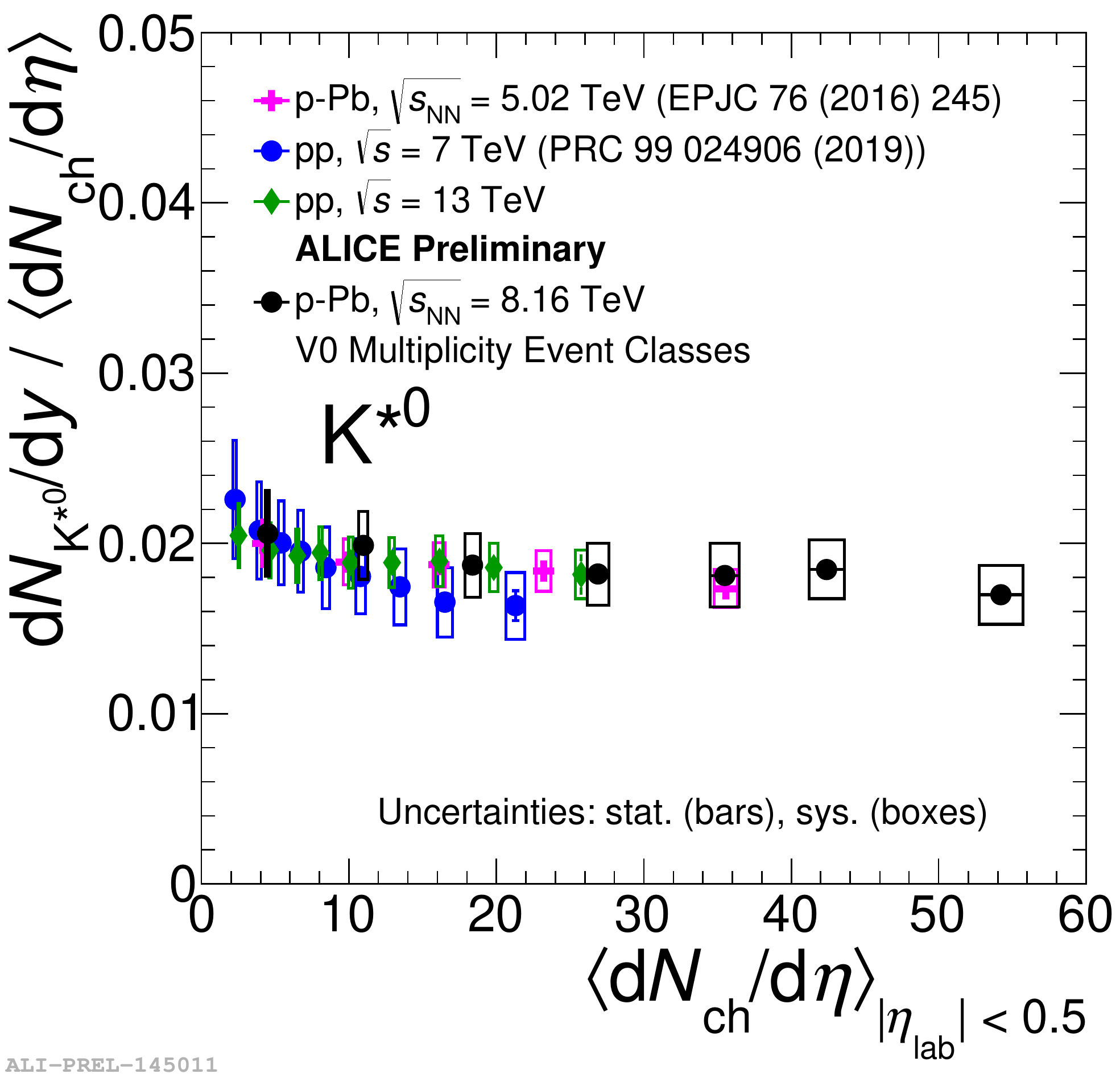}
\includegraphics[scale=0.23]{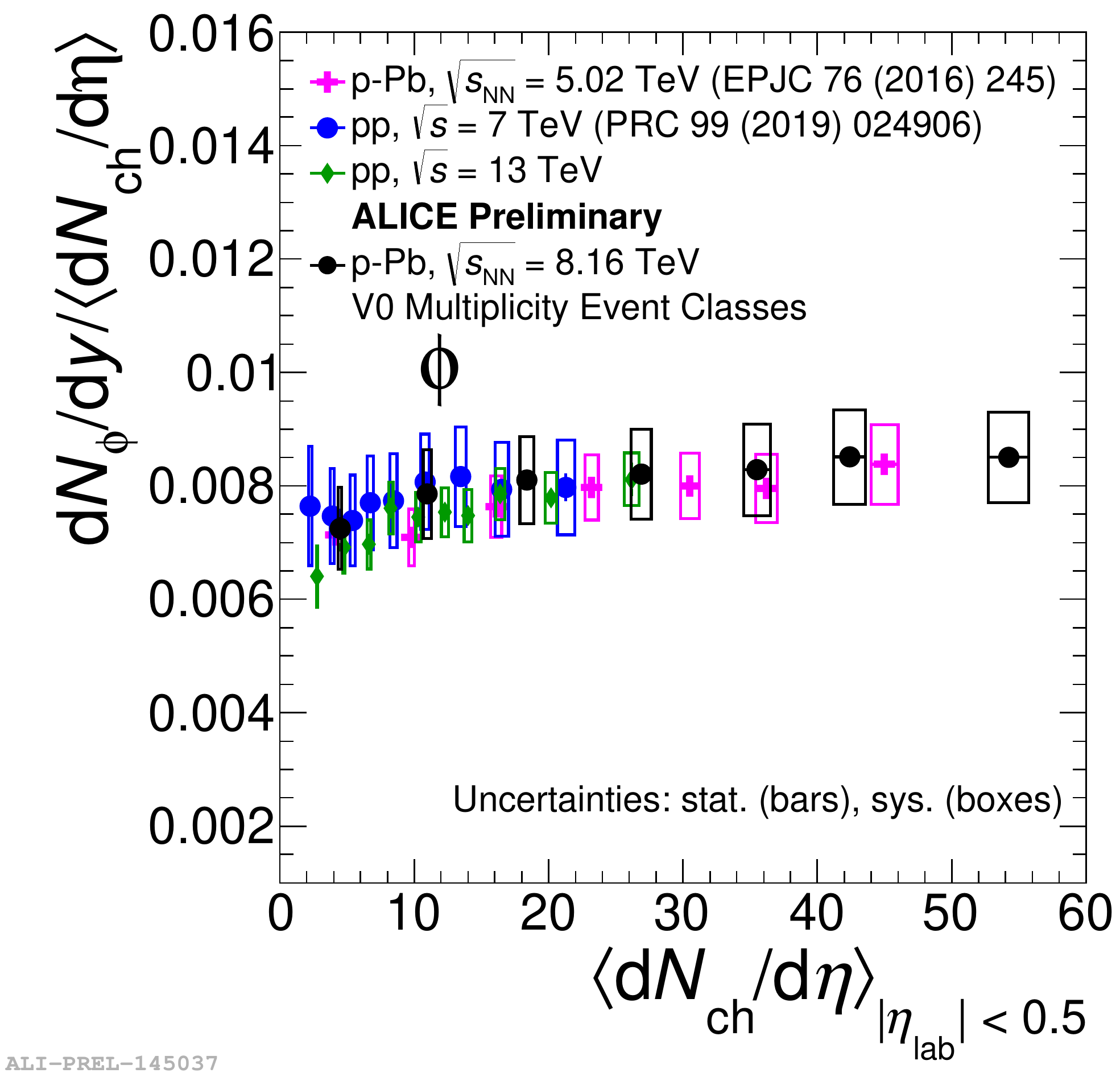}
}
\vspace{-0.3cm}
\end{minipage}\hfill
\begin{minipage}[c]{0.3\textwidth}
\caption{(Color online)  Integrated yields of K$^{*0}$ (left panel)  and $\phi$ (right panel) normalized to 
$\langle\mathrm{d}\it N_{\mathrm{ch}}/\mathrm{d}\eta\rangle$ in pp collisions (at $\sqrt{s}$ = 7 and 13 TeV) and  p--Pb 
collisions (at \snn~= 5.02 and 8.16 TeV) for different multiplicity classes. The bars and the boxes represent the statistical 
and systematic errors, respectively. }
\label{QM:dndy_nch}
\end{minipage}
\end{figure}

\begin{figure}[ht!]
\begin{minipage}[c]{0.65\textwidth}
\vspace{-0.3 cm}
\centerline{
\includegraphics[scale=0.30]{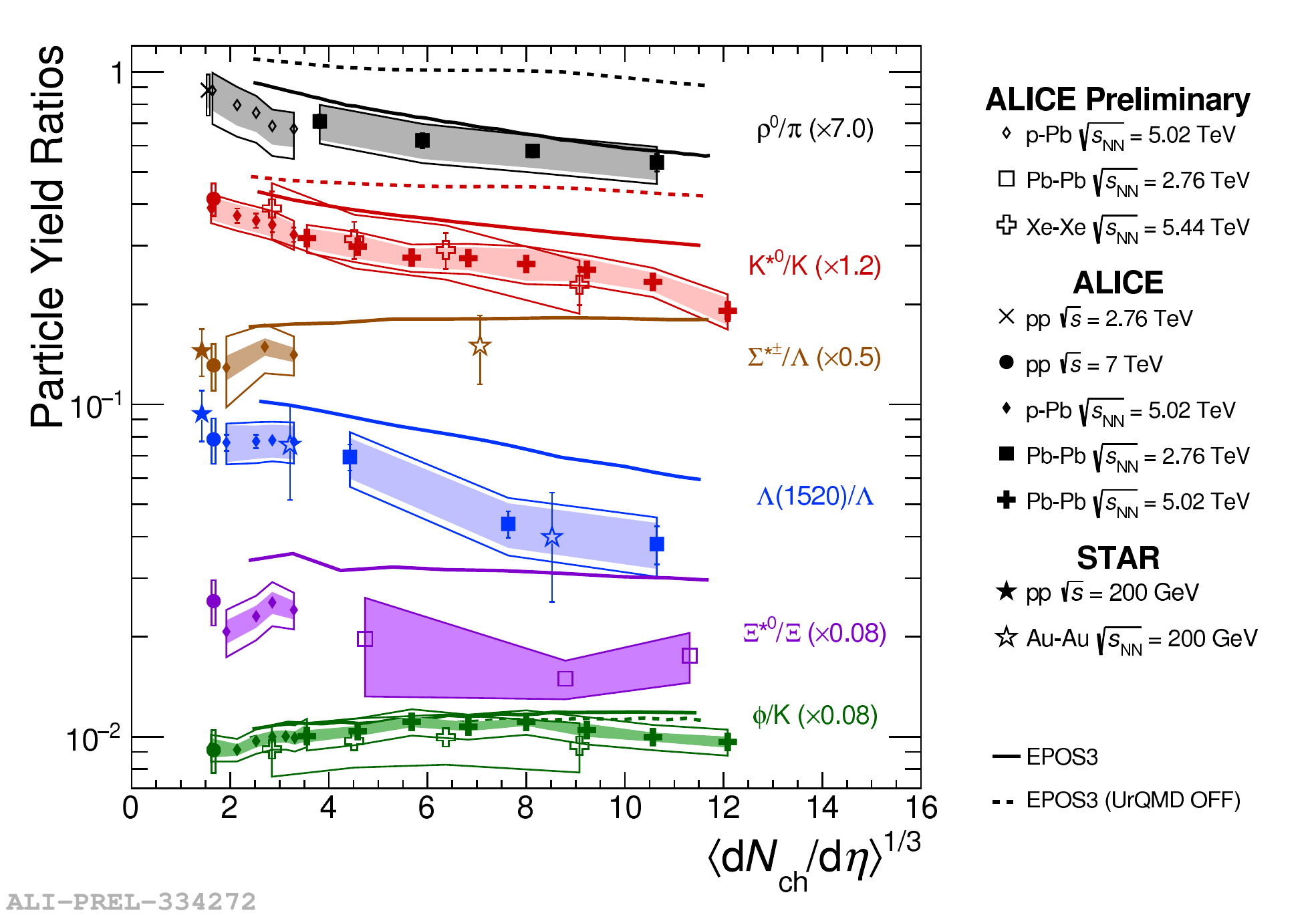}
\includegraphics[scale=0.25]{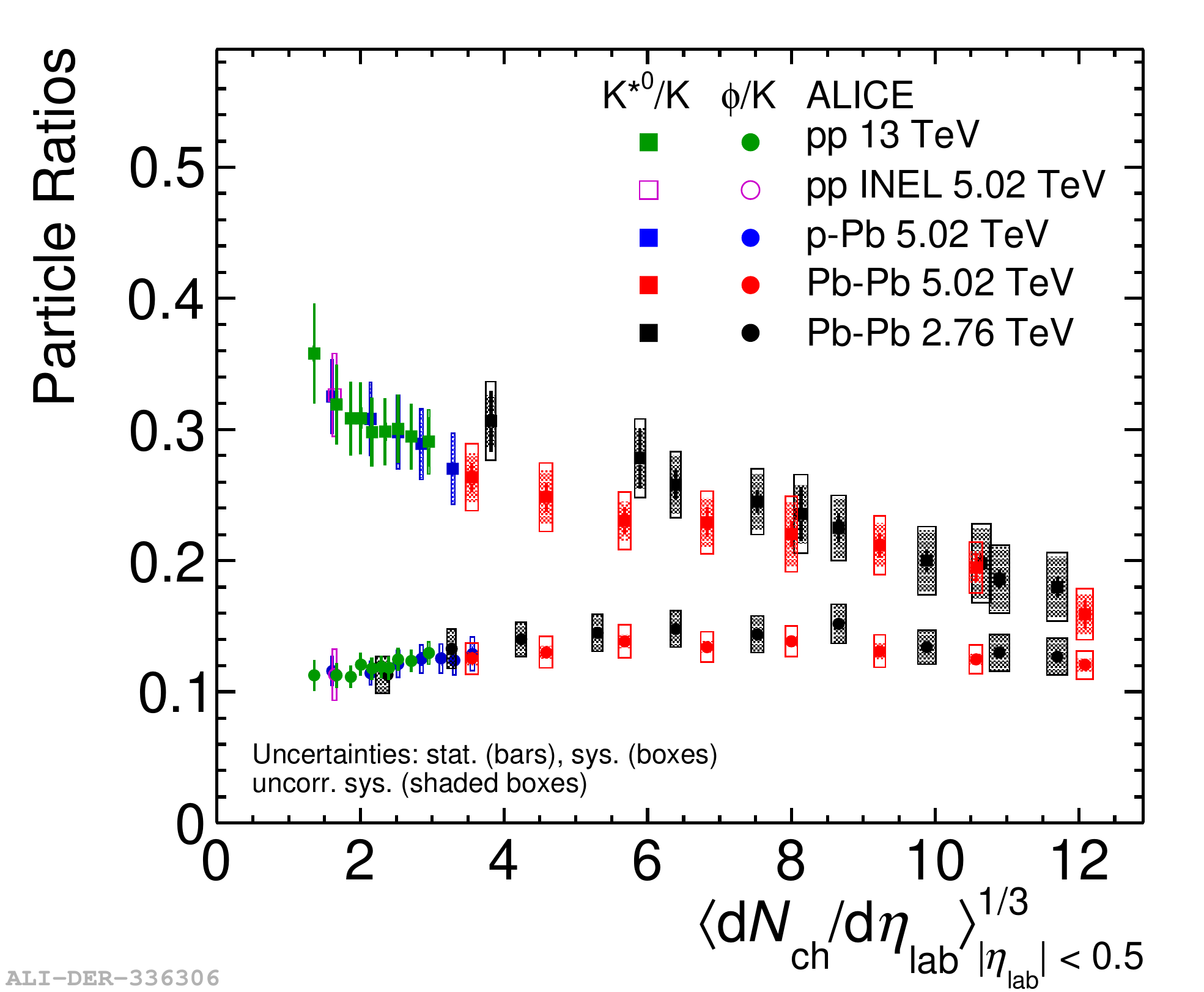}
}
\vspace{-0.3cm}
\end{minipage}\hfill
\begin{minipage}[c]{0.3\textwidth}
\caption{(Color online) Resonance yield ratio to long-lived hadrons  for pp, p--Pb, Xe--Xe, and Pb--Pb collisions, with comparisons 
to EPOS3 predictions and STAR data as a function of charged particle multiplicity ($\langle\mathrm{d}\it N_{\mathrm{ch}}/\mathrm{d}\eta\rangle$). 
(Left panel) The bars and the lines represent the statistical and systematic errors, respectively. (Right panel) The bars and the boxes represent
 the statistical and systematic errors, respectively. }
\label{QM:resonance_ratio}
\end{minipage}
\end{figure}

\begin{figure}[ht!]
\begin{minipage}[c]{0.65\textwidth}
\vspace{-0.3 cm}
\centerline{
\includegraphics[scale=0.24]{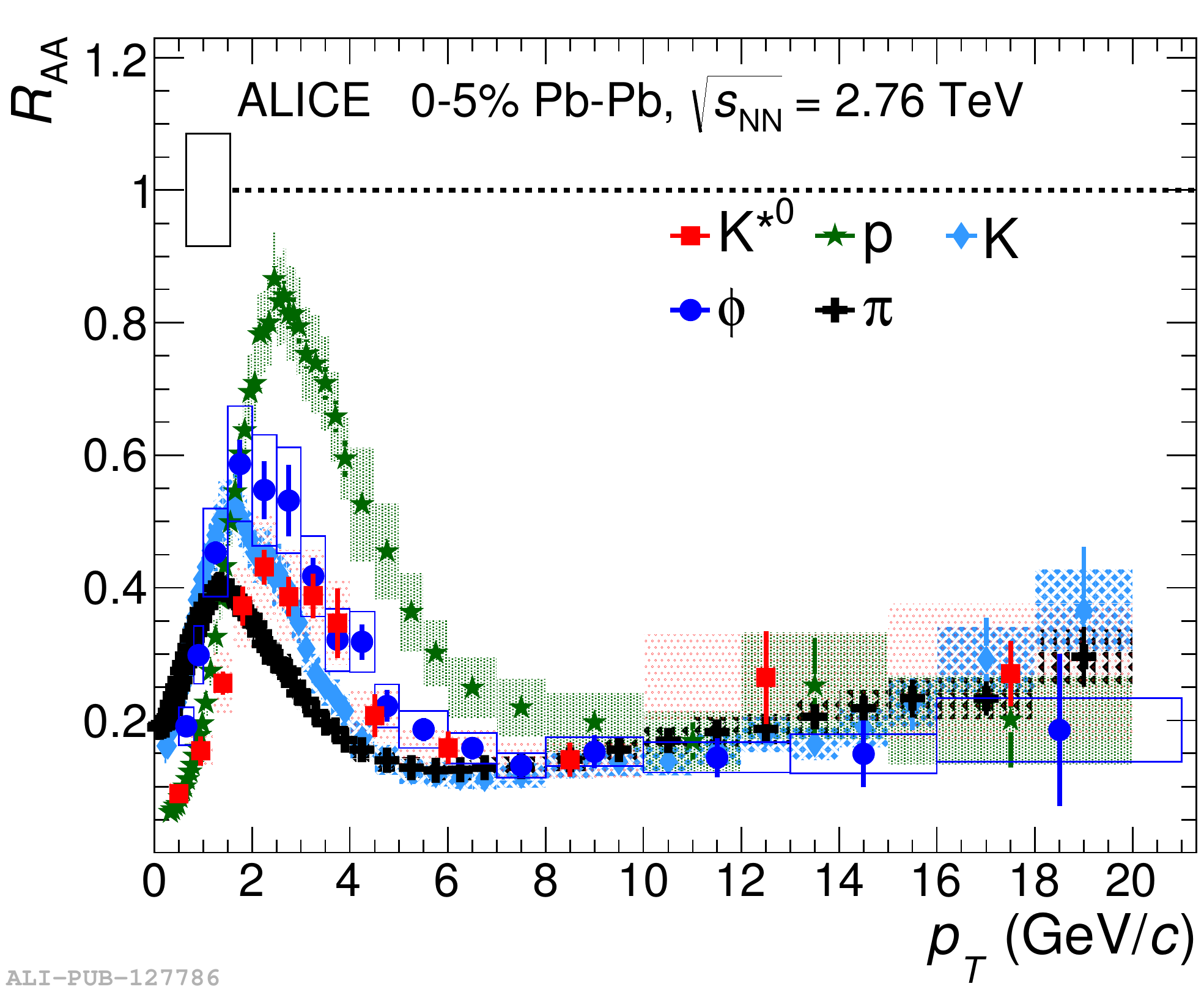}
}
\vspace{-0.8cm}
\end{minipage}\hfill
\begin{minipage}[c]{0.3\textwidth}
\caption{(Color online) The nuclear modification factor $R_{\rm{AA}}$, as a function of $p_{\mathrm{T}}$ for $\rm{K}^{*0}$ and $\phi$ mesons in 
$0-5\%$ centrality in Pb--Pb collisions at \snn~=~2.76 TeV.}
\label{QM:raa}
\end{minipage}
\end{figure}

The production of  $\rho(770)^{0}$, K$^{*}(892)$, $\phi(1020)$, $\Sigma(1385)^{\pm}$, $\Lambda(1520)$, and
$\Xi(1530)^{0}$ in different multiplicity classes in pp, p--Pb, Xe--Xe and Pb--Pb collisions in the energy range $\sqrt{s_{\rm NN}}$ = 2.76-13 
TeV have been measured in  a wide range of transverse momentum. The  $p_\mathrm{T}$-integrated  hadron yields ($\mathrm{d}\it{N}/\mathrm{d}\it{y}$) and 
mean $p_\mathrm{T}$ ($\langle p_{\mathrm{T}} \rangle$)  are determined by integrating the $p_\mathrm{T}$ spectra in the measured range and by using 
a   fit function (L\'{e}vy-Tsallis or Blast Wave) to extrapolate the yields in the unmeasured \pt~region for each multiplicity event class. 
Figure \ref{QM:meanpt} shows the comparison of  $\langle p_{\mathrm{T}} \rangle$ of $\mathrm{K}^{*0}$, $\phi$  and p as a function of charged particle multiplicity 
($\langle\mathrm{d}\it N_{\mathrm{ch}}/\mathrm{d}\eta\rangle$)  in pp collisions at $\sqrt{s}$ = 7 and 13 TeV, p--Pb at $\sqrt{s}$ = 5.02 and Pb--Pb at $\sqrt{s}$ = 2.76 TeV \cite{Acharya:2018orn}. A similar increase in $\langle p_{\mathrm{T}} \rangle$  with the charged particle multiplicity is observed for pp collisions at \snn~ = 7 and 13 TeV. 
A clear mass ordering of $\langle p_{\mathrm{T}} \rangle$ is observed in central Pb--Pb collisions, where $\mathrm{K}^{*0}$, 
$\phi$  and p with similar masses have similar $\langle p_{\mathrm{T}} \rangle$ values, which is expected from the hydrodynamic expansion 
of the system \cite{hydro}. However, the mass ordering of $\langle p_{\mathrm{T}} \rangle$ breaks down for the peripheral Pb--Pb and smaller collision systems.
The normalized integrated yields of K$^{*0}$ and $\phi$ to $\langle\mathrm{d}\it N_{\mathrm{ch}}/\mathrm{d}\eta\rangle$ in pp collisions at \snn~ = ~7 and 13 TeV, and  p--Pb 
collisions at \snn~=~ 5.02 and 8.16 TeV as a function of charged particle multiplicity are shown in Fig. \ref{QM:dndy_nch}. For both  K$^{*0}$ and $\phi$, it is observed that the yields normalised to $\langle\mathrm{d}\it N_{\mathrm{ch}}/\mathrm{d}\eta\rangle$ have similar values at same charged particle multiplicity and are independent of collision energy and system.
The ratio of resonance yields to the yields of long-lived hadrons as a function of charged particle multiplicity for $\rho^0/\pi$, $\rm{K}^{*0}/\rm{K}$, 
  $\Sigma^{*\pm}/\Lambda$, $\Lambda^{*}/\Lambda$, $\Xi^{*0}/\Xi$ and $\phi/\rm{K}$ in pp, p--Pb, Xe--Xe, and Pb--Pb collisions are shown 
  in Fig. \ref{QM:resonance_ratio}. Significant suppression is observed for $\rho^0/\pi$, $\rm{K}^{*0}/\rm{K}$, and $\Lambda^{*}/\Lambda$ with
  charged particle multiplicity in central Pb--Pb collisions. This is qualitatively explained by the EPOS generator with UrQMD, which attributes the suppression to 
  the decay daughters in the hadronic phase of the collision \cite{Knospe:2015nva,Acharya:2019qge}. For pp collisions, the decreasing trend of $\rm{K}^{*0}/\rm{K}$ with
  charged particle multiplicity hints at the possible presence of a hadronic phase with non-zero lifetime in  small collision systems \cite{Acharya:2019bli}.
\begin{figure}[ht!]
\begin{minipage}[c]{0.65\textwidth}
\vspace{-0.3 cm}
\centerline{
\includegraphics[scale=0.21]{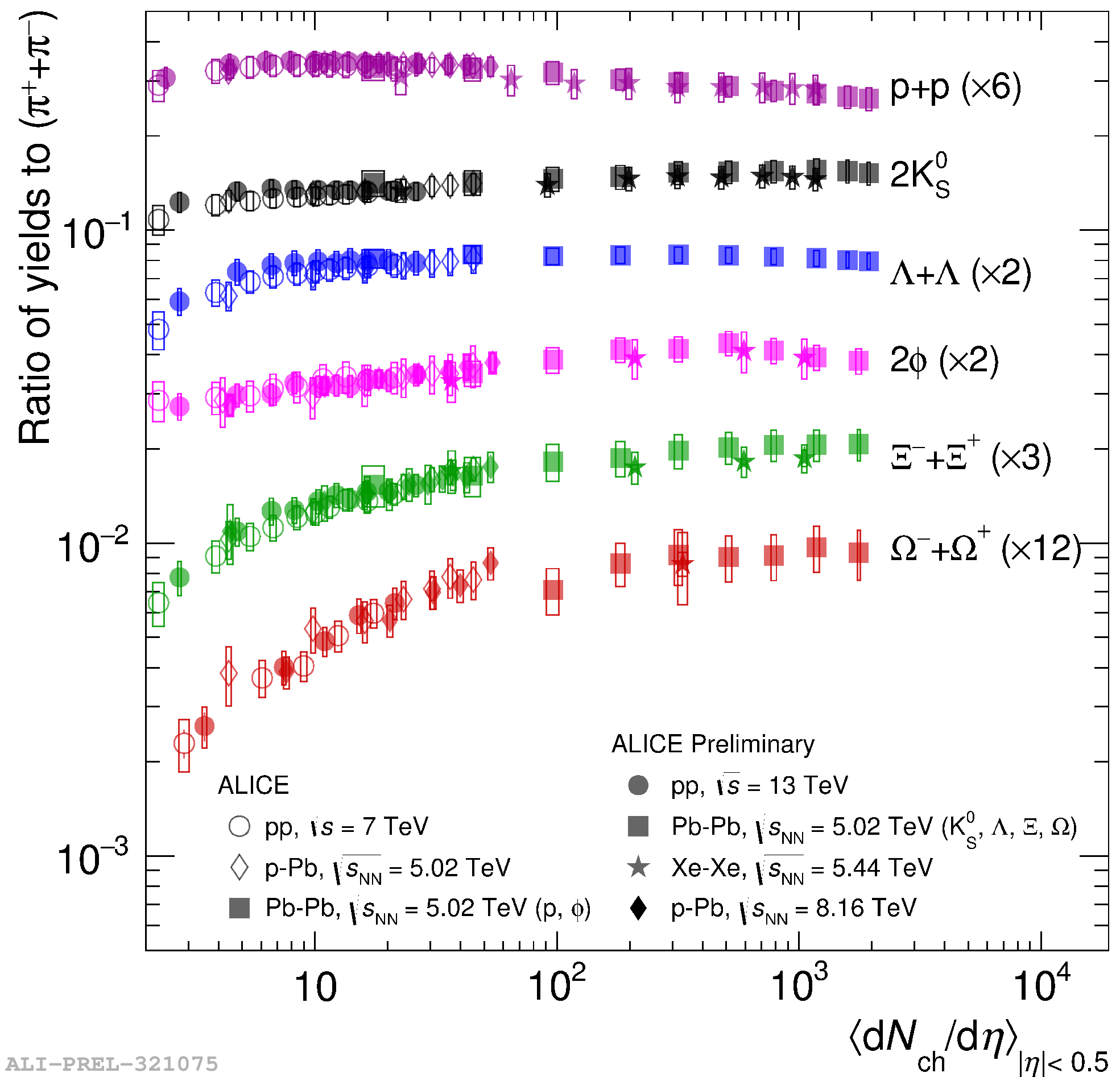}
\includegraphics[scale=0.23]{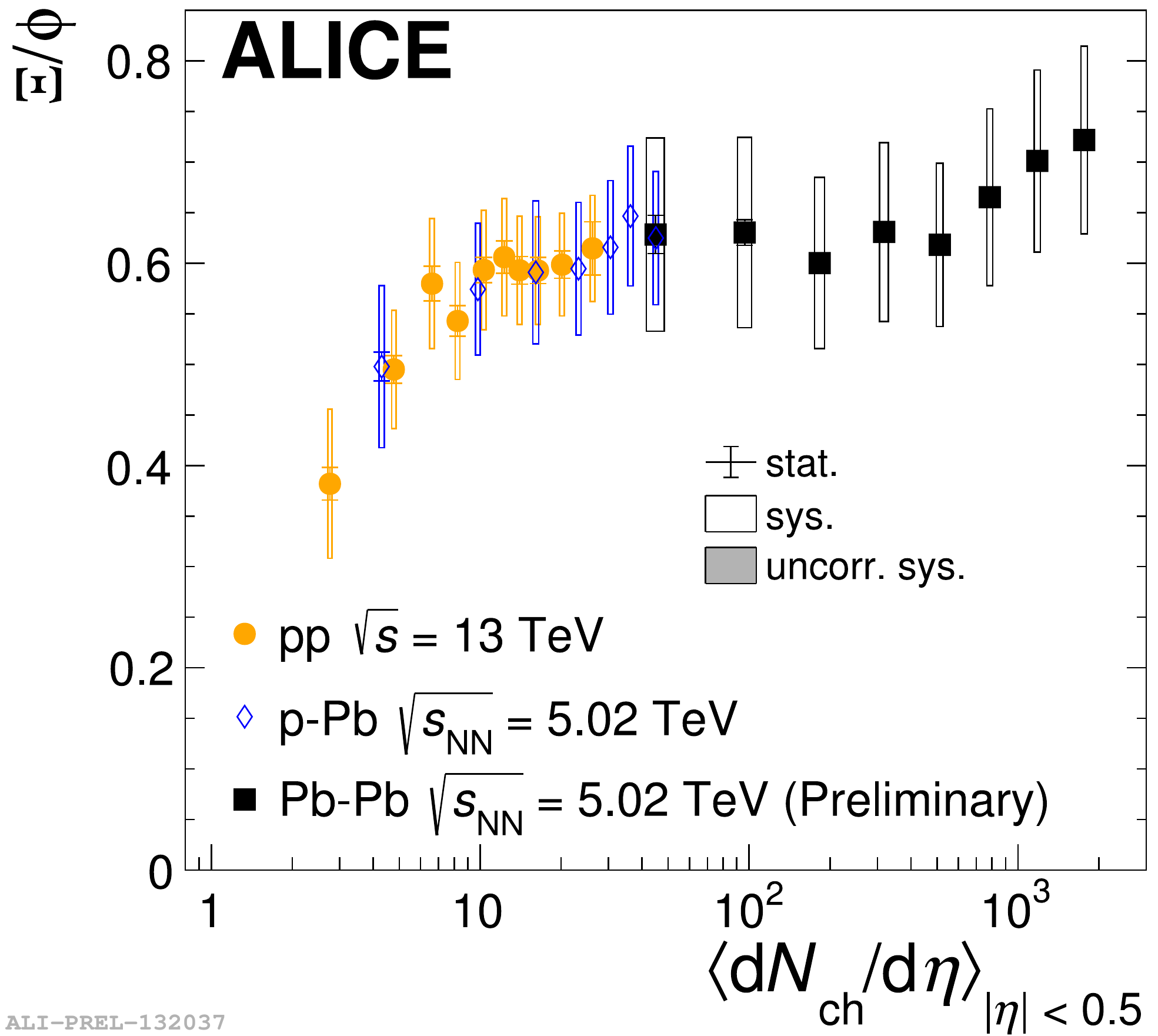}
}
\vspace{-0.3cm}
\end{minipage}\hfill
\begin{minipage}[c]{0.3\textwidth}
\caption{(Color online) Left panel: Ratios of the integrated yields, (p+$\bar{\mathrm{p}}$)/($\pi^{+}+\pi^{-}$), 2K$_\mathrm{S}^{0}$/($\pi^{+}+\pi^{-}$), 
    ($\Lambda+\bar{\Lambda}$)/($\pi^{+}+\pi^{-}$), $2\phi$/($\pi^{+}+\pi^{-}$),  ($\Xi^{-}+\bar{\Xi^{+}}$)/($\pi^{+}+\pi^{-}$), 
    ($\Omega+\bar{\Omega}$)/($\pi^{+}+\pi^{-}$), measured as a function of the charged particle density in pp 
    (at $\sqrt{s}$ = 7 and 13 TeV), p--Pb (at \snn~= 5.02 and 8.16 TeV), Xe--Xe (at \snn~=5.44) and Pb--Pb (at \snn~= 5.02 TeV) collisions. Right panel: 
    Ratios of the integrated yields, $\Xi$/$\phi$ in pp collisions at $\sqrt{s}$ = 13 TeV, p--Pb and Pb--Pb 
    collisions at \snn~= 5.02 TeV. The bars and the boxes represent the statistical and systematic error, respectively.}
\label{QM:ratio_pions}
\end{minipage}
\end{figure}
The ratios of hadron yields to pions as a function of $\langle\mathrm{d}\it N_{\mathrm{ch}}/\mathrm{d}\eta\rangle$ in pp at 
$\sqrt{s}$ = 7 and 13 TeV, p--Pb at \snn~= 5.02 and 8.16 TeV, Xe--Xe at \snn~=5.44 TeV, and Pb-Pb at \snn~= 5.02 TeV are shown 
in Fig.~\ref{QM:ratio_pions} \cite{Adam:2015vsf,Abelev:2013haa} (Left panel). We observe a smooth evolution of particle ratios with multiplicity in pp, p--Pb, Xe--Xe and 
Pb--Pb collisions at \snn~= 2.76 -13 TeV: for a given charged particle multiplicity strangeness production is independent of collision system 
and energy. The $\phi$ meson with net strangeness zero, is not subjected to the canonical suppression while the production of hadrons with 
open strangeness may be canonically suppressed. However, it is rather difficult to describe enhancement of $\phi$ meson production in a framework that 
involves canonical suppression \cite{Acharya:2018orn}.  In models that use core-corona \cite {Werner:2013tya} and rope-hadronization pictures \cite{Biro:1984cf}, the evolution of $\phi$ meson yields with charged particle multiplicity is similar to the behaviour for particles of open strangeness, making it possible for the measurement to provide explanations for  enhancement in small systems.
The $\phi$ meson production relative to pions increases with charged particle multiplicity as shown in Fig.  \ref{QM:ratio_pions} and also a mild enhancement for $\phi/\rm{K}$ ratios for small collision systems.
For the $\Xi/\phi$ ratio shown in Fig \ref{QM:ratio_pions} (Right panel) we see a flat or mild increasing trend 
for a wide range of multiplicity classes. These results suggests that the $\phi$-meson behaves as a particle with an effective strangeness of 1-2. 
Figure \ref{QM:raa} shows the nuclear modification factor as a function of $p_{\mathrm{T}}$ for $\rm{K}^{*0}$ and $\phi$ mesons in 
$0-5\%$ centrality in Pb--Pb collisions at \snn~= 2.76 TeV \cite{Adam:2015kca,Adam:2017zbf}. Similar suppression is observed for  different 
light flavour hadrons  above $p_{\mathrm{T}}$ $\simeq$ 10 GeV/$c$. Mass ordering is observed for $R_{\rm{AA}}$ at intermediate 
$p_{\mathrm{T}}$ among mesons, which may be attributed to the effect of radial flow.

\section{Summary}
\label{QM:sum}
ALICE has extensive results on production of resonances with lifetimes 1.3-46.4 fm/$c$ in different collision systems. The particle yields have  similar values for a given charged particle multiplicity: hadrochemistry is driven by event multiplicity rather than by colliding system and collision energy.
Mass ordering of $\langle p_{\mathrm{T}} \rangle$ is observed in central Pb--Pb collisions, which is expected from the hydrodynamic expansion of the system. However, this mass ordering breaks 
down for the peripheral Pb--Pb and smaller collision systems. Suppression of short-lived resonances, $\rho^0$, $\rm{K}^{*0}$, and  $\Lambda^{*}$ ($\tau$ ~$ <$ 13 fm/$c$) in most central A--A collisions, whereas no suppression is observed for the $\phi$ ($\tau$  = 46.4 fm/$c$) meson. The suppression in A--A collisions is attributed to the dominance of rescattering in the hadronic phase.  A hint of suppression for $\rm{K}^{*0}/\rm{K}$ in high multiplicity pp and p--Pb collisions suggests the presence of re-scattering effects in high multiplicity pp and p--Pb collisions. The $\phi/\rm{K}$ ratio shows mild increasing trend with multiplicity in pp collisions whereas  $\Xi/\phi$ shows an increasing trend over a wide range of multiplicity. These results together suggest that $\phi$ meson behaves as a particle with an effective strangeness between  1 and 2. Suppression of $R_{\rm{AA}}$   for  different light flavour hadrons  including resonances above $p_{\mathrm{T}}$ $\simeq$ 10 GeV/$c$ is similar, which indicates that parton energy loss at high  $p_{\mathrm{T}}$ is species independent.

\end{document}